\documentstyle[preprint,aps,graphicx]{revtex}
\def\mycomm#1{\hfill\break\strut\kern-3em{\tt ====> #1}\hfill\break}
\def\mycommNL#1{\strut\kern-3em{\tt ====> #1}\hfill\break}

\catcode`\@=11 
\def\lsim{\mathrel{\mathpalette\@versim<}}
\def\gsim{\mathrel{\mathpalette\@versim>}}
\def\@versim#1#2{\vcenter{\offinterlineskip
        \ialign{$\m@th#1\hfil##\hfil$\crcr#2\crcr\sim\crcr } }}
\catcode`\@=12 

\def\sbar{\hbox{$\bar s$}}

\def\thetap{\hbox{$\Theta^+$}}

\def\eqref#1{(\ref{#1})}

\makeatletter
\def\hlinewd#1{\noalign{\ifnum0=`}\fi
\hrule \@height #1 \futurelet \reserved@a\@xhline}
\def\hwhiteline{\noalign
{\ifnum0=`}\fi\hrule
\@height 0pt\vskip 1.0ex\futurelet \reserved@a\@xhline}
\makeatother
\def\gray{\special{ps: 0.40 setgray}}
\def\black{\special{ps: 0.0 setgray}}

\newcommand{\mydraft}{
\newcount\timecount
\newcount\hours \newcount\minutes  \newcount\temp \newcount\pmhours

\hours = \time
\divide\hours by 60
\temp = \hours
\multiply\temp by 60
\minutes = \time
\advance\minutes by -\temp
\def\hour{\the\hours}
\def\minute{\ifnum\minutes<10 0\the\minutes
    \else\the\minutes\fi}
\def\clock{
\ifnum\hours=0 12:\minute\ AM
\else\ifnum\hours<12 \hour:\minute\ AM
\else\ifnum\hours=12 12:\minute\ PM
    \else\ifnum\hours>12
     \pmhours=\hours
     \advance\pmhours by -12
     \the\pmhours:\minute\ PM
     \fi
    \fi
\fi
\fi
}
\def\fullclock{\hour:\minute}
\begin{centering}
\gray
\font\Hugett  =cmtt12 scaled\magstep4
\hbox{\Hugett Draft:\today,\clock}
\black
\end{centering}
\vskip -1.7cm
$\phantom{a}$
} 

\def\beq#1{\begin{equation} \label{#1}}
\def\eeq{\end{equation}}
\def\bra#1{\left\langle #1\right\vert}
\def\ket#1{\left\vert #1\right\rangle}

\newskip\humongous \humongous=0pt plus 1000pt minus 1000pt

\newif\ifdtup

\begin{document}
{\tighten
 \preprint {\vbox{
  \hbox{$\phantom{aaa}$}
  \vskip-1.5cm
\hbox{Cavendish-HEP-04/03}
\hbox{TAUP 2759-04} 
\hbox{WIS/03/04-JAN-DPP}
\hbox{ANL-HEP-PR-04-02}
}}

\title{The narrow width of the $\Theta^+$ - a possible explanation}

\author{Marek Karliner\,$^{a,b}$\thanks{e-mail: \tt marek@proton.tau.ac.il}
\\
and
\\
Harry J. Lipkin\,$^{b,c}$\thanks{e-mail: \tt
ftlipkin@clever.weizmann.ac.il} }
\address{ \vbox{\vskip 0.truecm}
$^a\;$Cavendish Laboratory\\
Cambridge University, England;\\
and\\
$^b\;$School of Physics and Astronomy \\
Raymond and Beverly Sackler Faculty of Exact Sciences \\
Tel Aviv University, Tel Aviv, Israel\\
\vbox{\vskip 0.0truecm}
$^c\;$Department of Particle Physics \\
Weizmann Institute of Science, Rehovot 76100, Israel \\
and\\
High Energy Physics Division, Argonne National Laboratory \\
Argonne, IL 60439-4815, USA\\
}
\maketitle
\begin{abstract}%

The narrow width of the exotic narrow baryon resonance $\Theta^+$  might be
explained  by mixing between the two nearly degenerate states that arise in
models with two diquarks and an antiquark. The only open $\Theta^+$ decay
channel is $KN$. When two states both coupled to a single dominant decay mode
are mixed by the loop diagram via this decay mode, diagonalization of the loop
diagram decouples one mass eigenstate from this decay mode as in some
treatments of the $\rho-\pi$ decay from the mixed  singlet-octet $\omega-\phi$
system,  the $K^* -\pi$ decay of the strange axial vector mesons and the $NK$
couplings of some baryons. This mechanism can explain the narrow width and weak
coupling of $\Theta^+ \rightarrow KN$ while allowing a relatively large
production cross section from $K^*$ exchange. Interesting tests are suggested
in $K^-p$ reactions where backward   kaon production must go by exotic baryon
exchange. 

\end{abstract}%

\vfill\eject
The recent experimental discovery of an exotic 5-quark $K N$ resonance
\cite{Kyoto,pentaexps}, 
\thetap\ with \hbox{$S={+}1$}, a mass of 1540 MeV, a very small width $\lsim 20$
MeV (possibly as little as $1{\div2}$~MeV~\cite{ThetaWidth}),
and a presumed quark configuration $uudd\sbar$  has given rise to a number of
theoretical models postulating diquark or triquark correlations
\cite{NewPenta,OlPenta,JW}.  However all have difficulty in explaining the 
narrow width.
There is also the problem of explaining comparatively large production cross
sections by diagrams involving kaon exchange with an $NK\Theta^+$ coupling
constant limited by the observed decay width.      

Models with two diquarks and an 
antiquark\cite{NewPenta,OlPenta,JW} have two different wave functions described
by different color and spin couplings of the three constituents. Any
simple model for the hyperfine interaction mixes these two
states\cite{jenmalt}.
 In similar situations with two-state systems coupled to a
single dominant decay mode, the mixing via loop diagrams has been shown to
create a decoupling of one of the eigenstates from this dominant decay mode. 
Some examples are		  
$\omega-\phi$
mixing\cite{katzozi}, the mixing of the strange axial vector
mesons\cite{axial}, and the ``ideal mixing"\cite{faiman} decoupling the $KN$ 
decay mode in p-wave decays of some negative parity strange baryons  
\cite{karlisg}. The suggestion that the narrow width of the $\Theta^+$ might be
due to a decoupling mechanism has also been made in a different 
context\cite{naftali}.

We apply this approach to the diquark  pentaquark models and show that the  
single dominant $KN$ decay mode is decoupled to a good approximation from one
of the two diquark-triquark states. The $KN$ decay mode includes the two decay
channels $K^+n$ and $K^0 p$ which have equal branching ratios in an isospin
conserving decay of an isoscalar state. Isospin breaking is negligible.
Therefore the two modes are both decoupled in   the same way and are considered
together in the following analysis of isoscalar $KN$ decays.

Let $\ket {\Theta_1}$ and $\ket {\Theta_2}$ denote any orthonormal basis for
the two diquark-diquark-antiquark states with different color and spin
couplings\cite{jenmalt}.

The eigenstates of the mass matrix will have the form 
\beq{ThetaSphi}
\ket \Theta_S \equiv 
\cos \phi \cdot \ket {\Theta_1} +
\sin \phi \cdot \ket {\Theta_2} 
\end{equation}
\beq{ThetaLphi}
\ket \Theta_L \equiv 
\sin \phi \cdot \ket {\Theta_1} -
\cos \phi \cdot \ket {\Theta_2} 
\end{equation}
Where $\phi$ is a mixing angle determined by the diagonalization of the mass
matrix.

Since each of these states can decay by quark rearrangement to the isoscalar
$KN$ final state, we define their decay transition matrix elements respectively
as $\bra{KN} T \ket{\Theta_1}$ and $\bra{KN} T \ket{\Theta_2}$.  We then find
that these two states can be mixed by a loop diagram
\beq{loop}
\Theta_i \rightarrow KN \rightarrow \Theta_j
\end{equation}
The contribution of this loop diagram to the mass matrix is 
\beq{loopmass}
M_{ij} = M_o\cdot \bra{\Theta_i} T \ket {KN}
\bra{KN} T \ket {\Theta_j}
\end{equation}
We first consider the approximation where 
$\ket {\Theta_1}$ and $\ket {\Theta_2}$ are degenerate and
the mass matrix is dominated by the
loop diagram contribution (\ref{loopmass}) and other contributions are
neglected. The mass eigenstates
(\ref{ThetaSphi}-\ref{ThetaLphi}) are:
\beq{ThetaS}
\ket \Theta_S = C\,\left[
\bra{KN} T \ket {\Theta_1} \cdot \ket {\Theta_1}
+
\bra{KN} T \ket {\Theta_2} \cdot \ket {\Theta_2}
\right]
\end{equation}
\beq{ThetaL}
\ket \Theta_L = C\,[
\bra{KN} T \ket {\Theta_2} \cdot \ket {\Theta_1}
-
\bra{KN} T \ket {\Theta_1} \cdot \ket {\Theta_2}
]
\end{equation}
where $C$ is a normalization
factor. Then  \beq{decThetaL} \bra{KN} T \ket \Theta_L =  \bra{KN} T \ket
{\Theta_1} \cdot \bra{KN} T \ket {\Theta_2} - \bra{KN} T \ket {\Theta_2} \cdot
\bra{KN} T \ket {\Theta_1}  = 0 \end{equation}

Thus in this approximation the state $\Theta_L$ is forbidden to decay into the 
$KN$ final state, while the $\Theta_S$ should have a normal hadronic width of
about 100 MeV and probably escape observation against the continuum
background. 

In contrast with their couplings to the $K N$ channel,
the couplings of the states $\ket {\Theta_1}$ and $\ket {\Theta_2}$  to $K^*N$
final states will not satisfy these relations and therefore both states $\ket
{\Theta_L}$ and $\ket {\Theta_S}$ can be produced without any suppression by
$K^*$ exchange. 

This model thus predicts a stronger production via $K^*$ exchanges than would
be obtained from kaon exchange with an $NK\Theta^+$ coupling limited by the
observed decay width.      

An interesting further test of this model would be in the baryon-exchange
$K^- p$ reactions where the kaon is observed going backward in the
center-of-mass system:
\beq{baryonex}
K^-p \rightarrow \bar K^o n; ~ ~ ~
K^- p\rightarrow  \bar K^{*o} n; ~ ~ ~ K^-p \rightarrow \bar K^o N^{*o};
~ ~ ~
K^- p\rightarrow  \bar K^{*o} N^{*o}
\end{equation}
where  $N^{*o}$ denotes any $I=1/2$ electrically neutral baryon resonance.

\vskip1cm
\begin{center}
\includegraphics[width=14.0em,clip=true,angle=90]{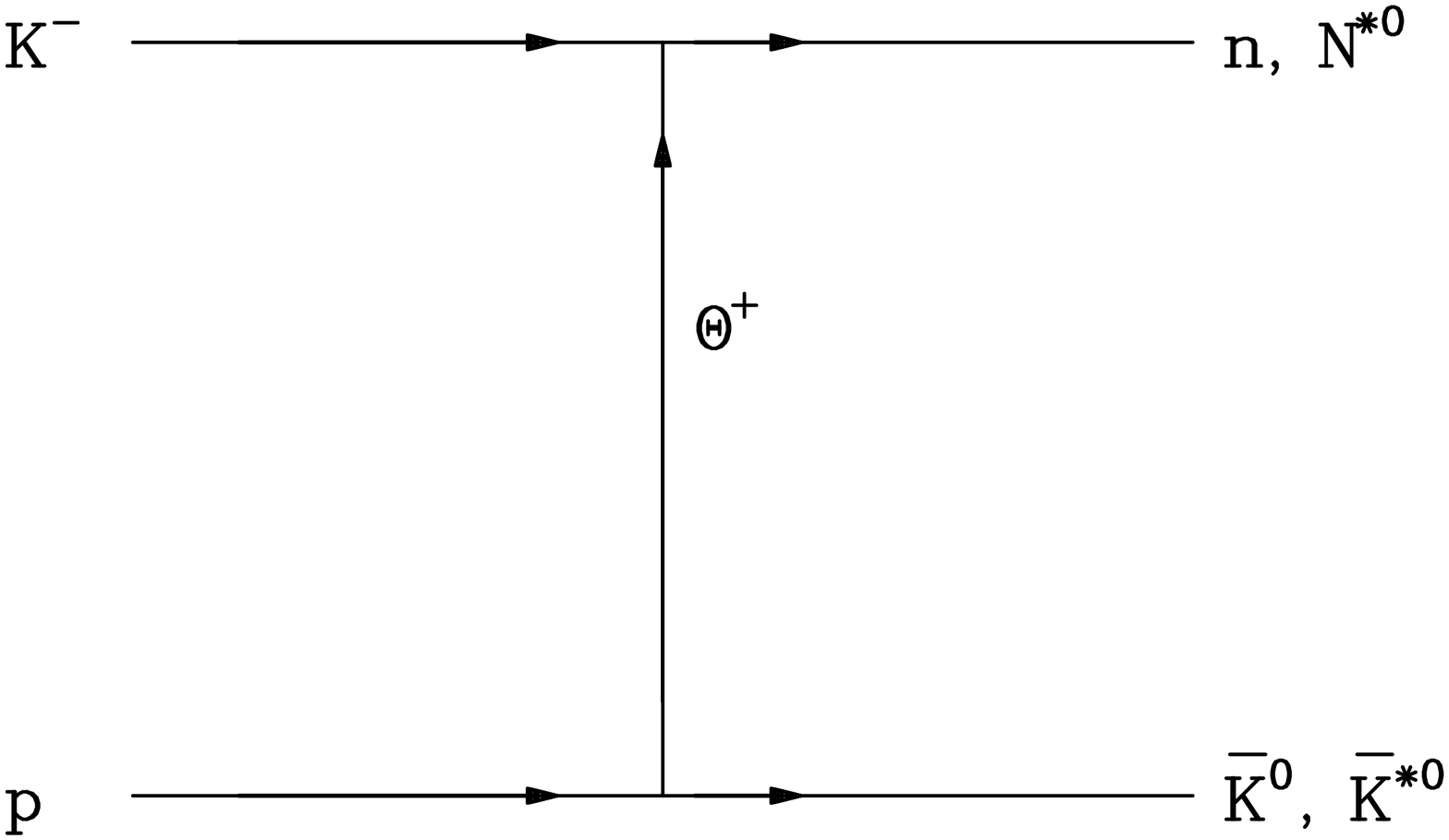}
\vbox{\small Fig. 1. Baryon-exchange diagram corresponding
to the reactions (\ref{baryonex}).}
\end{center}
\vskip1cm

These reactions shown in Fig. 1 can only proceed via the exchange of an
exotic
positive-strangeness baryon. But if the $\Theta^+$  couples only weakly to
$KN$, the $K^- \Theta^+ \rightarrow n$ and $p \rightarrow \bar K^o \Theta^+$
vertices
are also weak by crossing and the $K^- p\rightarrow  \bar K^{*o} N^{*o}$
reaction should be much stronger than the other three which require a
$\Theta^+KN$ coupling.

The $\Theta^+K\Delta$ coupling is forbidden by
isospin if the $\Theta^+$ is an isoscalar. Therefore the presence
of the $\Delta $ in these baryon exchange reactions is a test for the
presence of exotic positive strangeness baryons with higher isospin.

A more precise calculation, not feasible at present, will consider other
contributions to the  mass matrix in addition to the loop diagram.  However,
we can show by a rough calculation how this mechanism can explain a suppression of the  width of the $\Theta^+$ in any
model where two states contribute to the decay and the other contributions to
the mass matrix are not much greater than the decay width of the broad state. 

In addition to the loop diagram  contribution to the mass matrix,  denoted by
$M_o$, let us add  a contribution  which splits the masses of the two states in
the absence of the loop diagram. This contribution is denoted by off-diagonal
matrix elements $M_1$ in the base where the 
$\ket{\Theta_L}$ and  $\ket {\Theta_S}$
states are diagonal. The mass matrix then becomes

\beq{massS}
\bra{\Theta_S} M \ket{\Theta_S} = M_o; ~ ~ ~ 
\bra{\Theta_S} M \ket{\Theta_L} =M_1  
\end{equation}
\beq{massL}
\bra{\Theta_L} M \ket{\Theta_S} = M_1; ~ ~ ~ 
\bra{\Theta_L} M \ket{\Theta_L} = 0 
\end{equation}

From the trace and the determinant of this matrix the mass eigenvalues 
denoted by $M_A$ and $M_B$ are
\beq{masseig1}
M_A = M_o + \xi; ~ ~ ~ M_B = -\xi ; ~ ~ ~ M_A-M_B=M_o + 2\xi 
\end{equation}
where
\beq{masseig2}
\xi \cdot (M_o + \xi) = M_1^2
\end{equation}

The masses are no longer degenerate in the absence of the loop diagram. The
splitting without the loop diagram is obtained by setting $M_o=0$ in 
eqs. (\ref{masseig1}- \ref{masseig2})
\beq{masssplit}
\Delta M = 2M_1
\end{equation}

The eigenstates of this mass matrix differ from the loop diagram eigenstates by
a mixing angle denoted by $\alpha$

\beq{ThetaA}
\ket {\Theta_A} \equiv 
\ket {\Theta_S} \cdot \cos \alpha +
\ket {\Theta_L} \cdot \sin \alpha
\end{equation}
\beq{ThetaB}
\ket {\Theta_B} \equiv 
\ket {\Theta_S} \cdot \sin \alpha -
\ket {\Theta_L} \cdot \cos \alpha
\end{equation}

where
\beq{massalpha}
\tan \alpha = {{\xi}\over{M_1}} = {{M_1}\over{M_o + \xi}}
= {{\Delta M}\over{2(M_o + \xi)}}
\end{equation}

The ratio of the $KN$ decay widths of the two states is thus
 
\beq{widthratio}
{{|\bra{KN} T \ket{\Theta_B}|^2}\over{|\bra{KN} T \ket{\Theta_A}|^2}} = 
\tan^2 \alpha = {{\xi^2}\over{M_1^2}} = {{M_1^2}\over{(M_o + \xi)^2}}
= {{\Delta M^2}\over{4(M_o + \xi)^2}}
\leq {{\Delta M^2}\over{4M_o^2}}
\end{equation}

We thus see that the narrow state $\ket{\Theta_B}$ is narrower than the normal
broad state $\ket{\Theta_A}$ if the mass splitting $\Delta M$ between the two 
states is smaller than the loop diagram contribution $M_o$ which gives the
width of the broad state. For example for a splitting $\Delta M$ = 50 MeV and
a broad width $M_o$ = 100 MeV, the width of the narrow state is suppressed by a
factor 16.

\section*{Acknowledgements}

The research of one of us (M.K.) was supported in part by a grant from the
United States-Israel Binational Science Foundation (BSF), Jerusalem and
by the Einstein Center for Theoretical Physics at the Weizmann Institute.
The research of one of us (H.J.L.) was supported in part by the U.S. Department
of Energy, Division of High Energy Physics, Contract W-31-109-ENG-38.
We benefited from discussions with A. Auerbach, D. Faiman and G. Karl who 
called our attention to refs \cite{naftali,faiman,karlisg}. 
%
\catcode`\@=11 
\def\references{
\ifpreprintsty \vskip 10ex
%
\hbox to\hsize{\hss \large \refname \hss }\else
\vskip 24pt \hrule width\hsize \relax \vskip 1.6cm \fi \list
{\@biblabel {\arabic {enumiv}}}
{\labelwidth \WidestRefLabelThusFar \labelsep 4pt \leftmargin \labelwidth
\advance \leftmargin \labelsep \ifdim \baselinestretch pt>1 pt
\parsep 4pt\relax \else \parsep 0pt\relax \fi \itemsep \parsep \usecounter
{enumiv}\let \p@enumiv \@empty \def \theenumiv {\arabic {enumiv}}}
\let \newblock \relax \sloppy
 \clubpenalty 4000\widowpenalty 4000 \sfcode `\.=1000\relax \ifpreprintsty
\else \small \fi}
\catcode`\@=12 

}


\begin{thebibliography}{99}

\bibitem{Kyoto}
T.~Nakano {\it et al.}  [LEPS Coll.],
Phys.\ Rev.\ Lett.\  {\bf 91}(2003)012002,
hep-ex/0301020.

\bibitem{pentaexps}
V.~V.~Barmin {\it et al.}  [DIANA Collaboration],
Phys.\ Atom.\ Nucl.\  {\bf 66} (2003) 1715
[Yad.\ Fiz.\  {\bf 66} (2003) 1763],
hep-ex/0304040;
S.~Stepanyan {\it et al.}  [CLAS Collaboration],
hep-ex/0307018.
%
J.~Barth {\it et al.}  [SAPHIR Collaboration],
hep-ex/0307083;
%
V.~Kubarovsky and S.~Stepanyan  and CLAS Collaboration,
hep-ex/0307088;
%
A.~E.~Asratyan, A.~G.~Dolgolenko and M.~A.~Kubantsev,
hep-ex/0309042.
%
V.~Kubarovsky et al., [CLAS Collaboration],
hep-ex/0311046;
A. Airapetian et al., [HERMES Collaboration],
arXiv:hep-ex/0312044;
S.~Chekanov, [ZEUS Collaboration],
{\tt http://www.desy.de/f/seminar/Chekanov.pdf}.

\bibitem{ThetaWidth}
S. Nussinov,  hep-ph/0307357;
R.~W.~Gothe and S.~Nussinov,
hep-ph/0308230;
%
R.~A.~Arndt, I.~I.~Strakovsky and R.~L.~Workman,
Phys.\ Rev.\ C {\bf 68} (2003) 042201,
nucl-th/0308012 and nucl-th/0311030;
%
J. Haidenbauer and G. Krein,  hep-ph/0309243;
%
R.~N.~Cahn and G.~H.~Trilling,
hep-ph/0311245;
%
A.~Casher and S.~Nussinov,
Phys.\ Lett.\ B {\bf 578} (2004) 124,
hep-ph/0309208.

\bibitem{NewPenta}
M. Karliner and H.J. Lipkin,
Phys.\ Lett.\ B {\bf 575} (2003) 249.

\bibitem{OlPenta}
M. Karliner and H.J. Lipkin,
[arXiv:hep-ph/0307243].

\bibitem{JW} R. L. Jaffe and F. Wilczek, 
hep-ph/0307341
\bibitem{jenmalt} Byron K. Jennings and Kim Maltman, hep-ph/0308286
\bibitem{katzozi}
A. Katz and H. J. Lipkin,
Phys.\ Lett. {\bf 7} (1963) 44
\bibitem{axial}
Harry J. Lipkin,
Phys.\ Lett.\ B {\bf 72} (1977) 249.
\bibitem{faiman}
D. Faiman,
Phys.\ Rev.\ D {\bf 15},  (1977) 854
\bibitem{karlisg}
Nathan Isgur and Gabriel Karl
Phys.\ Lett. {\bf 74} (1978) 35
\bibitem{naftali} N. Auerbach and V. Zelevinsky, nucl-th/0310029 


\end{thebibliography}
\end{document}